\documentclass[preprintnumbers,10pt,nofootinbib]{revtex4}

\usepackage{amsmath,latexsym,amssymb,amsfonts}
\usepackage[dvips]{graphicx,color}
\usepackage{bm}
\usepackage{epstopdf} 

\begin{document}
\title{\textbf{Fuzzy Euclidean wormholes in anti-de Sitter space}}

\author{
\textsc{Subeom Kang$^{a,b}$}\footnote{{\tt ksb527{}@{}kaist.ac.kr}} and \textsc{Dong-han Yeom$^{c,d,e}$}\footnote{{\tt innocent.yeom{}@{}gmail.com}}
}

\affiliation{
$^{a}$Department of Physics, KAIST, Daejeon 34141, Republic of Korea\\
$^{b}$Center for Theoretical Physics of the Universe, IBS, Daejeon 34051, Republic of Korea\\
$^{c}$Leung Center for Cosmology and Particle Astrophysics, National Taiwan University, Taipei 10617, Taiwan\\
$^{d}$Asia Pacific Center for Theoretical Physics, Pohang 37673, Republic of Korea\\
$^{e}$Department of Physics, POSTECH, Pohang 37673, Republic of Korea
}

\begin{abstract}
This paper is devoted to an investigation of Euclidean wormholes made by fuzzy instantons. We investigate the Euclidean path integral in anti-de Sitter space. In Einstein gravity, we introduce a scalar field with a potential. Because of the analyticity, there is a contribution of complex-valued instantons, so-called fuzzy instantons. If we have a massless scalar field, then we obtain Euclidean wormholes, where the probabilities become smaller and smaller as the size of the throat becomes larger and larger. If we introduce a non-trivial potential, then in order to obtain a non-zero tunneling rate, we need to tune the shape of the potential. With the $O(4)$ symmetry, after the analytic continuation to the Lorentzian time, the wormhole throat should expand to infinity. However, by adding mass, one may obtain an instant wormhole that should eventually collapse to the event horizon. The existence of Euclidean wormholes is related to the stability or unitarity issues of anti-de Sitter space. We are not conclusive yet, but we carefully comment on these physical problems.
\end{abstract}

\maketitle

\newpage

\tableofcontents

\section{Introduction}

Understanding gravity as a quantum theory is the ultimate goal of modern theoretical physics. There are several approaches that can be taken to investigate quantum gravity. The first approach, the so-called \textit{covariant approach} \cite{DeWitt:1967ub}, is to introduce a Lagrangian and calculate several observables of quantum fluctuations using the path integral formalism, This approach has the problem of non-renormalizability, but by extending to the string theory, one may obtain a finite probability. However, in order to study non-perturbative effects directly, another alternative approach is needed. The second approach is to use a Hamiltonian, in the so-called \textit{canonical approach} \cite{DeWitt:1967yk}. In this approach, we can introduce the Wheeler-DeWitt equation for the wave function of the Universe. One modern extension toward this direction is called loop quantum gravity, although it introduces lots of difficulties and many issues that need to be clarified.

In quantum gravity, there is a technical subtlety in the following sense. In order to understand fully quantum gravitational phenomena, e.g., to understand a singularity, we need to study non-perturbative effects. However, in the covariant approach, what we can understand technically well are the perturbative calculations using field theoretical techniques. While holography or duality is useful, since one can transform a non-perturbative regime into a perturbative regime as its dual \cite{Maldacena:1997re}, this does not lead to a direct understanding of the non-perturbative phenomena. On the other hand, with the canonical approach, there is no well-defined field theoretical basis and we need to go beyond the traditional field theoretical techniques to obtain a correct understanding of quantum gravitational phenomena. Then following question becomes this: is there any theoretical way to understand non-perturbative quantum gravitational phenomena, at least in an approximate way, using field theoretical tools?

The Euclidean path integral approach \cite{Gibbons:1994cg}, which is Euclidean since it introduces complex time, is a good bridge between these two approaches; one can study non-perturbative phenomena using well-established field theoretical tools, at least in an approximate way. Of course, the Euclidean path integral approach is not the perfect way to understand quantum gravity because of several weak points, including the fact that the Euclidean action may not be bounded from below. On the other hand, at the approximate level, it can give reliable results, including black hole entropy \cite{Gibbons:1976ue}, Hawking radiation \cite{Hartle:1976tp}, the probability distribution of thermal fluctuations \cite{Linde:1993xx} that can be interpreted as thermal instantons \cite{Hawking:1981fz}, etc.

The main approximation technique of the Euclidean path integral approach is to use instantons for the steepest-descent approximation \cite{Hartle:1983ai}. Because of the analyticity, instantons can be complex-valued. However, as with the path integral between two hypersurfaces, each hypersurface should be real-valued. Therefore, we need to impose a reality condition at the boundary, the so-called classicality condition \cite{Hartle:2007gi}. The authors of \cite{Hartle:2007gi} have referred to these complex-valued but would-be classicalized instantons as \textit{fuzzy instantons}. One can investigate various interesting phenomena using fuzzy instantons in the context of inflationary cosmology or modified gravity \cite{Hwang:2011mp}.

When we use a scalar field, the imaginary counterpart of the scalar field looks a ghost field, i.e., the imaginary part of the scalar field has the wrong sign for its kinetic term. This is apparently problematic, but in the Euclidean path integral we can allow this as long as we can set a suitable boundary condition (classicality condition), since these imaginary values naturally appear during the approximation of the wave function. Using this imaginary part of the scalar field, one can see interesting phenomena with many aspects \cite{Chen:2015ria}. One such example is the Euclidean wormhole \cite{Hawking:1988ae}.

In this paper, we investigate Euclidean wormholes using fuzzy instantons, especially in anti-de Sitter space. Fuzzy Euclidean wormholes in de Sitter space were the focus of the authors' previous paper \cite{Chen:2016ask}. Several authors have investigated Euclidean wormholes in the anti-de Sitter background in the string theory \cite{Giddings:1987cg,Maldacena:2004rf,ArkaniHamed:2007js}. Usually, this is related to the unitarity or the stability issue of anti-de Sitter space, but again this can be controversial. 

In this paper, we will show that the existence of the Euclidean wormholes is very generic by introducing fuzzy instantons (hence, it does not sensitively depend on a specific model of the string theory). On the other hand, we need to introduce a classicality constraint to ensure a sensible physical interpretation. This raises interesting issues. In the paper, we present this problem and open up interesting questions for further investigations.

This paper is organized as follows. In SEC.~\ref{sec:euc}, we discuss the mathematical formalism for the fuzzy instantons and the suitable initial conditions of instantons for Euclidean wormholes. In SEC.~\ref{sec:pro}, we discuss the details of the Euclidean wormhole solutions, including their probabilities as well as Lorentzian dynamics. In SEC.~\ref{sec:dyn}, we generalize the wormhole dynamics by using the thin-shell approximation. From this approximation, we can see more of the generic behaviors of wormholes. Finally, in SEC.~\ref{sec:con}, we summarize this paper and discuss possible future topics that need to be clarified.

\section{\label{sec:euc}Euclidean wormholes using fuzzy instantons}

In this section, we first discuss the Euclidean path integral approach, the use of instantons, and the notion of classicality. In particular, we discuss the details of $O(4)$-symmetric instantons. In order to obtain a Euclidean wormhole, we need to establish a suitable initial condition of the instanton. Here, we discuss the details of these issues. 

\subsection{The Euclidean path integral}

In the Euclidean path integral approach, the Euclidean propagator between two hypersurfaces from the in-state $| h_{ab}^{\mathrm{i}}, \chi^{\mathrm{i}} \rangle$ to the out-state $| h_{ab}^{\mathrm{f}}, \chi^{\mathrm{f}} \rangle$ is represented by the following path integral \cite{Hartle:1983ai}:
\begin{eqnarray}
\langle h_{ab}^{\mathrm{f}}, \chi^{\mathrm{f}} | h_{ab}^{\mathrm{i}}, \chi^{\mathrm{i}} \rangle = \int \mathcal{D}g_{\mu\nu}\mathcal{D}\phi \;\; e^{-S_{\mathrm{E}}[g_{\mu\nu},\phi]},
\end{eqnarray}
where $S_{\mathrm{E}}$ is the Euclidean action which is a functional of the metric $g_{\mu\nu}$ and a (scalar) matter field $\phi$. Here, $h_{ab}$ is a three-surface and $\chi$ is a field value on $h_{ab}$. This path integral should sum over all field combinations that connect from $h_{ab}^{\mathrm{i}}$ and $\chi^{\mathrm{i}}$ to $h_{ab}^{\mathrm{f}}$ and $\chi^{\mathrm{f}}$. This will be approximated by steepest-descents, i.e., by summing Euclidean on-shell histories \cite{Hartle:1983ai}. These on-shell solutions are called instantons.

In general, due to the Wick-rotation of time, we require that all functions be complex-valued. Hartle, Hawking and Hertog designated these complex-valued instantons \textit{fuzzy instantons} \cite{Hartle:2007gi}. However, not all fuzzy instantons are relevant to a tunneling process. After a long Lorentzian time, at the observed hypersurfaces (both of the in-state or the out-state), all fields should be real-valued. This is related to the notion of classicality \cite{Hartle:2007gi}. If we approximately write the wave function using the steepest-descent approximation as
\begin{equation}
\Psi[q_{I}] \simeq A[q_{I}] e^{i S[q_{I}]},
\end{equation}
where $q_{I}$ are canonical variables with labels $I=1,2,3, ...$, then \textit{classicality} means that
\begin{equation} \label{eqn:classicality}
\left|\nabla_I A\left[q_{I}\right]\right|\ll \left|\nabla_I S\left[q_{I}\right]\right|,
\end{equation}
for all $I$. By imposing this condition, this history satisfies the semi-classical Hamilton-Jacobi equation. This history is classical in the sense that the given probability through the history does not vary rapidly\footnote{Of course, in order to satisfy the true steepest-descent approximation, one needs to check whether it is indeed the peak of the wave function or not, e.g., see \cite{Callan:1977pt}, and there is a possibility that an on-shell solution may not represent a steepest-descent, e.g., see \cite{Coleman:1987rm}. However, this goes beyond the scope of this paper and we postpone the discussion for a future work.}. When we solve on-shell Euclidean equations, although we introduce complex-valued functions, such complex-valued fields should approach real-valued functions after the Wick rotation and after a long Lorentzian time \cite{Hwang:2011mp}.

Therefore, we can conclude that in the Euclidean path integral approach, the use of complex-valued instantons is allowed as long as the instantons are well-controlled by the classicality condition. Keeping this in mind, we can begin to investigate fuzzy instantons in more detail.

\subsection{Model of fuzzy Euclidean instantons}

We consider a model with a scalar field and its potential. The Euclidean action is given by
\begin{eqnarray}
S_{\mathrm{E}} = - \int \sqrt{+g}dx^{4} \left[ \frac{R}{16\pi} - \frac{1}{2} \left(\nabla \Phi\right)^{2} - V(\Phi) \right],
\end{eqnarray}
where $R$ is the Ricci scalar, $\Phi$ is a scalar field, and
\begin{eqnarray}
V(\Phi) = V_{0} \left( - 1 - \frac{1}{2} \mu^{2} \Phi^{2} + \lambda \Phi^{4} \right).
\end{eqnarray}
The metric ansatz for $O(4)$-symmetry is
\begin{eqnarray}
ds^{2}_{\mathrm{E}} = \frac{1}{V_{0}} \left(d\tau^{2} + a^{2}(\tau) d\Omega_{3}^{2}\right).
\end{eqnarray}
Then, Euclidean equations of motion are as follows:
\begin{eqnarray}
\dot{a}^{2} - 1 - \frac{8\pi a^{2}}{3} \left( \frac{\dot{\Phi}^{2}}{2} - \frac{V(\Phi)}{V_{0}} \right) &=& 0,\\
\ddot{\Phi} + 3 \frac{\dot{a}}{a} \dot{\Phi} - \frac{V'(\Phi)}{V_{0}} &=& 0, \\
\frac{\ddot{a}}{a} + \frac{8\pi}{3}\left( \dot{\Phi}^{2} + \frac{V(\Phi)}{V_{0}} \right) &=& 0.
\end{eqnarray}
Since the potential is normalized by $V_{0}$, without any change in dynamics, we can choose $V_{0} = 1$. The one exception is the value of the Euclidean action. The Euclidean action will be scaled: $S_{\mathrm{E}}/V_{0}$, where $S_{\mathrm{E}}$ is the result for the case of $V_{0} = 1$.

In this paper, we consider fuzzy Euclidean wormholes, meaning that complex-valued fields are allowed for the bulk region, while we have to impose classicality at the boundary. In this regard, we consider a purely imaginary scalar field: $\Phi \rightarrow i \phi$. Then the effective equations of motions are as follows:
\begin{eqnarray}
\dot{a}^{2} - 1 - \frac{8\pi a^{2}}{3} \left( \frac{- \dot{\phi}^{2}}{2} - U(\phi) \right) &=& 0,\\
\ddot{\phi} + 3 \frac{\dot{a}}{a} \dot{\phi} - F(\phi) &=& 0, \\
\frac{\ddot{a}}{a} + \frac{8\pi}{3}\left( - \dot{\phi}^{2} + U(\phi) \right) &=& 0,
\end{eqnarray}
where
\begin{eqnarray}
U(\phi) &=& - 1 + \frac{1}{2} \mu^{2} \phi^{2} + \lambda \phi^{4},\\
F(\phi) &=& - \mu^{2} \phi - 4 \lambda \phi^{3}.
\end{eqnarray}
Here, $F$ is not a simple derivation of $U$; instead, we need to choose this form in order to maintain the correct equations of motion. Throughout this paper, we consider $\mu^{2} \geq 0$ and $\lambda \geq 0$. 

\subsection{Initial conditions}

In order to choose a consistent initial condition, we assign as follows:
\begin{eqnarray}
a(0) &=& a_{0},\\
\dot{a}(0) &=& 0,\\
\phi(0) &=& \phi_{0},\\
\dot{\phi}(0) &=& \dot{\phi}_{0},
\end{eqnarray}
where $a_{0}$ is a solution of
\begin{eqnarray}
1 + \frac{8\pi a_{0}^{2}}{3} \left( - \frac{\dot{\phi}_{0}^{2}}{2} - U(\phi_{0}) \right) = 0.
\end{eqnarray}
Here, we have chosen $\phi_{0} = 0$ for the symmetry of the solution. Therefore, $\dot{\phi}_{0}$ is the unique free parameter to characterize a solution.

\begin{figure}
\begin{center}
\includegraphics[scale=0.3]{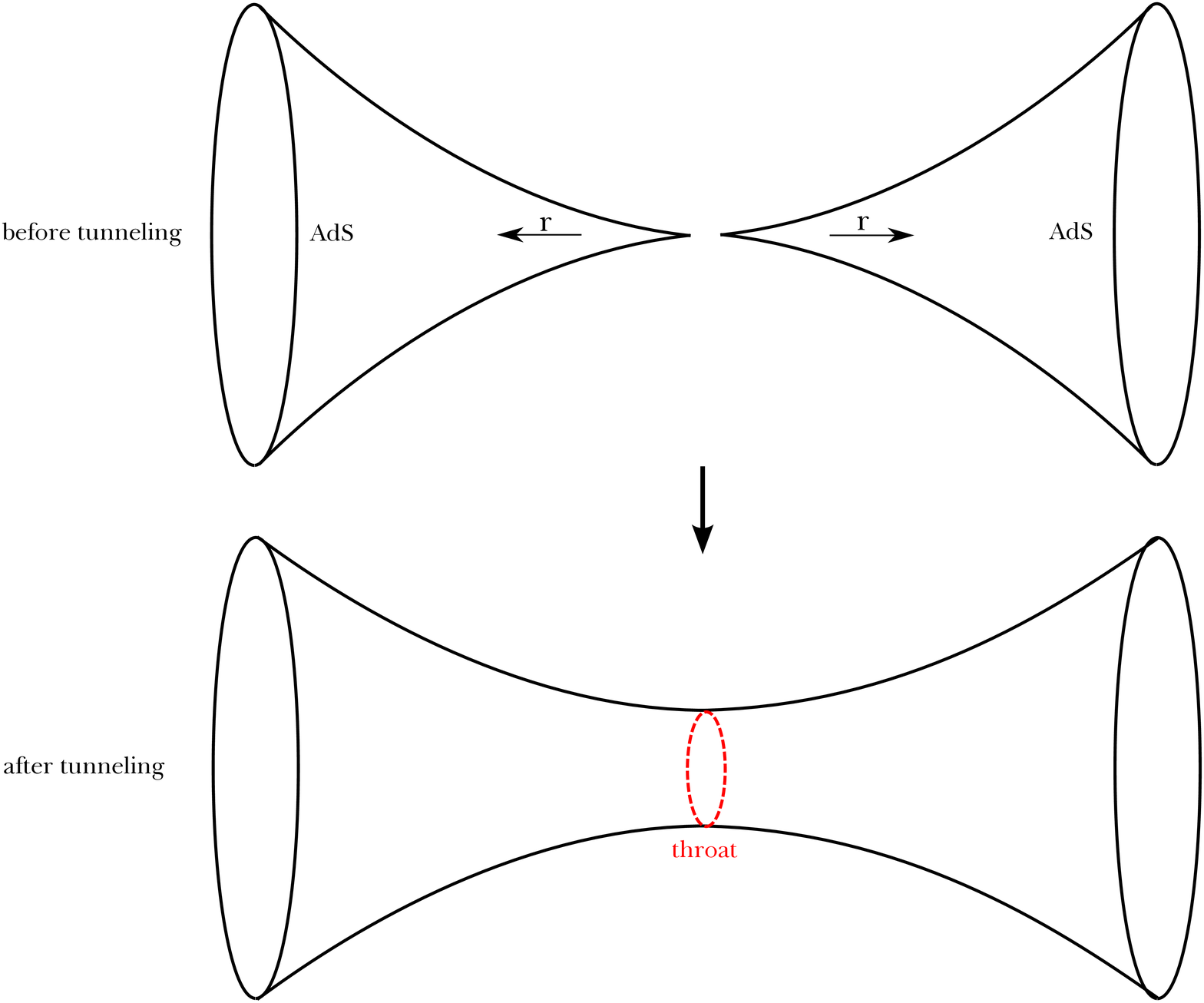}
\caption{\label{fig:concept}Conceptual picture of tunneling in a Euclidean wormhole. Before tunneling, two anti-de Sitter spaces are disconnected. After tunneling, two asymptotic anti-de Sitter spaces are connected at the throat.}
\end{center}
\end{figure}

\subsection{Conceptual picture}

Using instantons, what we want to study is the nucleation of a Euclidean wormhole. The initial condition is a hypersurface of two complete anti-de Sitter spaces. The upper part of FIG.~\ref{fig:concept} illustrates this. If we describe the anti-de Sitter space as a static coordinate
\begin{eqnarray}
ds^{2} = - \left( 1 + \frac{r^{2}}{\ell^{2}} \right) dt^{2} + \left( 1 + \frac{r^{2}}{\ell^{2}} \right)^{-1} dr^{2} + r^{2} d\Omega^{2}
\end{eqnarray}
with a suitable $\ell$, then two separate anti-de Sitter spaces can be denoted by two directions of $r$, where they are separated at $r=0$.

After tunneling, what we encounter is a hypersurface that connects two asymptotic anti-de Sitter spaces (the lower part of FIG.~\ref{fig:concept}). Before tunneling, the two anti-de Sitter spaces were separated, but after tunneling, the two surfaces are connected by a throat with a non-zero areal radius. In order to provide a reasonable anti-de Sitter limit to this wormhole hypersurface, we impose the classicality condition at both boundaries ($r \rightarrow \infty$).

Now, there  is usually a problem with distinguishing the interpretation of the Lorentzian signatures from the Euclidean signatures. In the $O(4)$-symmetric metric ansatz\footnote{There are interesting issues about boundary conditions of \cite{Henneaux:1985tv}, but in our example, a complexified scalar field effectively violates the null energy condition for a quantum regime. Quantum effects can violate energy conditions and can introduce a timelike wormhole in anti-de Sitter space, for example, see \cite{Harlow:2018tqv}, and hence it is not so surprising to see a connection of two anti-de Sitter boundaries.}, we can present the angle parameters as follows: $d\Omega_{3}^{2} = d\chi + \sin^{2} \chi d\Omega_{2}^{2}$. What we will use is the analytic continuation of $\chi = \pi/2 + iT$ as is usual with inhomogeneous instantons \cite{Coleman:1980aw}. (There would be an alternative way to interpret this as a homogeneous tunneling process \cite{Maldacena:2004rf}, but this is less interesting in the anti-de Sitter background. Perhaps, there are more interesting applications in the de Sitter background; see \cite{Chen:2016ask,Bramberger:2017cgf}.) Then, at the nucleation point $T=0$, we obtain $a(\tau) = r$, where $a$ is the scale factor of the instanton and $r$ is the areal radius of the static coordinate. Of course, the total metric will depend on the time $T$ and a detailed description of the wormhole after nucleation will be provided in the following sections. In any case, at this stage, it is enough to say that the behavior of $a$ is closely related to the behavior of $r$. In order to obtain a wormhole structure of $r$, what we will study is how to obtain a bouncing solution of $a$.

In this regard, in order to calculate the tunneling rate, we will need to subtract two Euclidean actions, where one is the action of the solution and the other is the action of \textit{two} anti-de Sitter spaces.

\section{\label{sec:pro}Properties of Euclidean wormholes}

In this section, we discuss details of the solution. Specifically, we will focus on finding a bouncing Euclidean solution for $a$. Of course, the existence of a bouncing solution is not enough. If the solution has a physical meaning, we then need to obtain a suitable probability. The probability should not be zero nor be too large; in other words, the subtracted Euclidean action should not be too large or too negative (if the subtracted Euclidean action is negative, then the tunneling process is exponentially enhanced). In this section, we will cover these various topics.

\subsection{Anti-de Sitter with a free scalar field}

The first example is the simplest case: $\mu = \lambda = 0$. In other words, if the potential is $V(\Phi) = -1$, then
\begin{eqnarray}
\frac{\ddot{\Phi}}{\dot{\Phi}} = - 3 \frac{\dot{a}}{a},
\end{eqnarray}
and hence
\begin{eqnarray}
\dot{\Phi} = \mathcal{A} a^{-3}
\end{eqnarray}
with a constant $\mathcal{A}$. The equation for $a$ becomes
\begin{eqnarray}
\dot{a}^{2} + V_{\mathrm{eff}}(a) &=& 0,\\
V_{\mathrm{eff}}(a) &=& - 1 - \frac{8\pi}{3} \left( \frac{\mathcal{A}^{2}}{2a^{4}} + a^{2} \right).
\end{eqnarray}
Hence, $V_{\mathrm{eff}}(a) < 0$ is the physically allowed region.

\begin{figure}
\begin{center}
\includegraphics[scale=0.85]{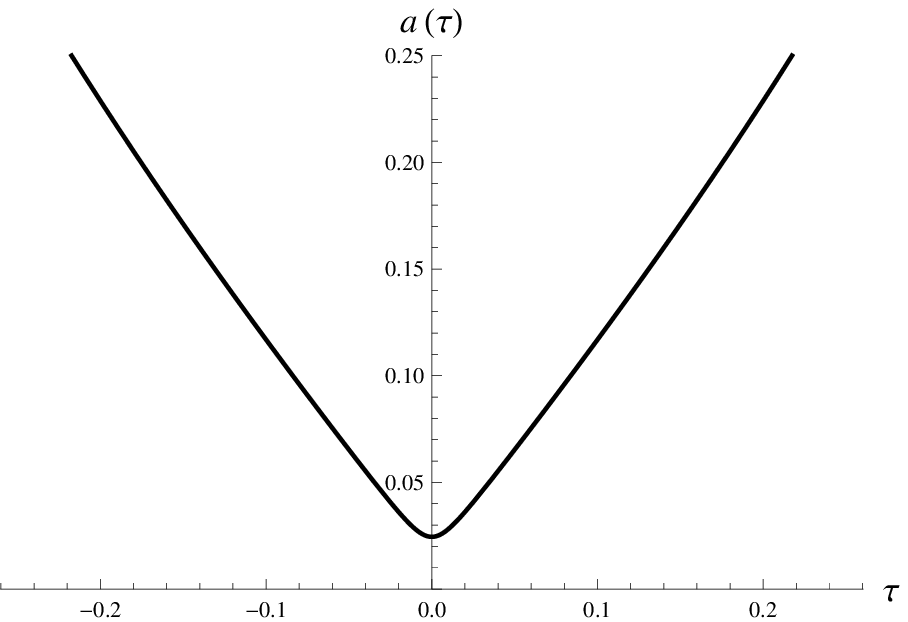}
\includegraphics[scale=0.85]{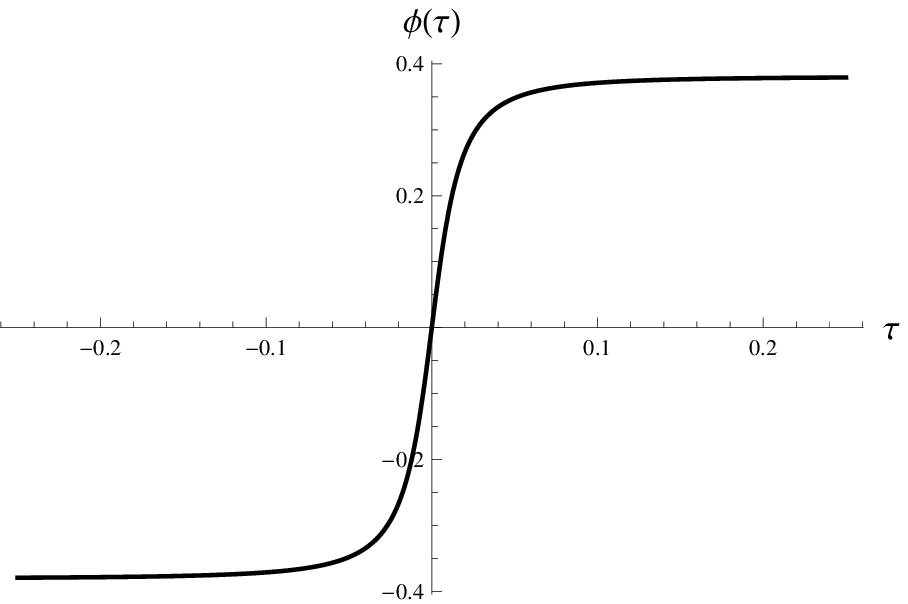}
\caption{\label{fig:sol0}$a(\tau)$ and $\phi(\tau) (= -i \Phi)$ for $\dot{\phi}_{0} = 20$ and $\mu = \lambda = 0$.}
\end{center}
\end{figure}

Assuming $\mathcal{A} = i \mathcal{B}$ with a real value $\mathcal{B}$, the effective potential becomes
\begin{eqnarray}
V_{\mathrm{eff}}(a) = - 1 + \left( \frac{a_{0}^{4}}{a^{4}} - \frac{a^{2}}{\ell^{2}} \right),
\end{eqnarray}
where $a_{0}=(4\pi \mathcal{B}^{2}/3)^{1/4}$ and $\ell = (3/8\pi)^{1/2}$. Then there can be a solution $a_{\mathrm{min}}$ that satisfies $V_{\mathrm{eff}}(a_{\mathrm{min}}) = 0$ and the solution is allowed for $a \geq a_{\mathrm{min}}$. Hence, $a_{\mathrm{min}}$ becomes the throat of the wormhole between two asymptotically anti-de Sitter spaces.

FIG.~\ref{fig:sol0} shows a numerical example of this. The field will stop asymptotically. Therefore, if there is no potential, one can trust the classicality at infinity.

Regarding the probability, the on-shell action where the scale factor moves from $a_{\mathrm{max}} (=\infty)$ through $a_{\mathrm{min}}$ to $a_{\mathrm{max}}$ again is
\begin{eqnarray}
S_{\mathrm{E}} &=& 4 \pi^{2} \int d\tau \left[ a^{3} V - \frac{3}{8\pi} a \right]\\
&=& - 3\pi \int_{a_{\mathrm{min}}}^{a_{\mathrm{max}}} da \frac{a \left(1 + \frac{a^{2}}{\ell^{2}} \right)}{\sqrt{1 - \left( \frac{a_{0}^{4}}{a^{4}} - \frac{a^{2}}{\ell^{2}} \right)}}.
\end{eqnarray}
The decay rate is $\Gamma \propto \exp -B$ and $B$ is the subtracted Euclidean action, where
\begin{eqnarray}
B = S_{\mathrm{E}} (\mathrm{final}) - S_{\mathrm{E}} (\mathrm{initial}).
\end{eqnarray}
As we mentioned before, the former term is the Euclidean action for the solution and the latter term is the Euclidean action for two anti-de Sitter spaces.

The subtracted action can be simplified as follows:
\begin{eqnarray}
B = - 3\pi \int_{a_{\mathrm{min}}}^{\infty} da \frac{a \left(1 + \frac{a^{2}}{\ell^{2}} \right)}{\sqrt{1 + \left( - \frac{a_{0}^{4}}{a^{4}} + \frac{a^{2}}{\ell^{2}} \right)}} + 3\pi \int_{0}^{\infty} da \frac{a \left(1 + \frac{a^{2}}{l^{2}} \right)}{\sqrt{1 + \frac{a^{2}}{\ell^{2}}}}.
\end{eqnarray}
In general, both terms are divergent, since the volumes of the anti-de Sitter spaces are infinite. However, by matching each $a$, the divergence can be regularized for some cases. In this flat potential case, for a large $a$ limit, this integration is expanded as
\begin{eqnarray}
B = 3\pi \int^{\infty} da \left( - \frac{a_{0}^{4} \ell}{2} \frac{1}{a^{4}} + \mathcal{O}\left( a^{-6} \right)\right) + (\mathrm{finite\;numbers}).
\end{eqnarray}
Therefore, we can be sure that the divergences will be canceled at infinity.

\begin{figure}
\begin{center}
\includegraphics[scale=0.85]{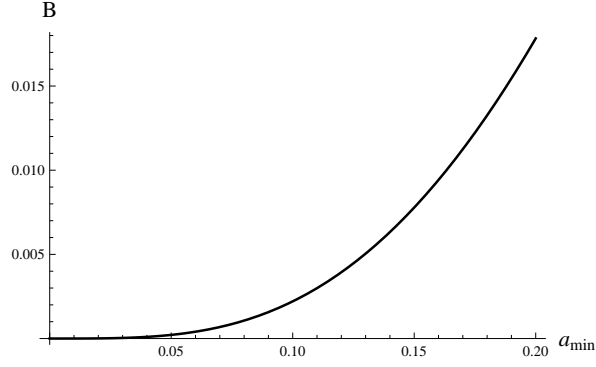}
\caption{\label{fig:probability}$B$ by varying $a_{\mathrm{min}}$ (assuming $V_{0}=1$).}
\end{center}
\end{figure}

As we can be sure of the convergence of the integration, we can calculate more details. We present this subtracted action as follows:
\begin{eqnarray}
B &=& \int_{a_{\mathrm{min}}}^{\infty} \mathcal{F}(a) da + 3\pi \int_{0}^{a_{\mathrm{min}}} a \sqrt{1 + \frac{a^{2}}{\ell^{2}}} da\\
&=& \int_{a_{\mathrm{min}}}^{\infty} \mathcal{F}(a) da + \pi \ell^{2} \left[ \left( 1 + \frac{a_{\mathrm{min}}^{2}}{\ell^{2}} \right)^{3/2} - 1 \right],
\end{eqnarray}
where
\begin{eqnarray}
\mathcal{F} = - 3\pi \frac{a \left(1 + \frac{a^{2}}{\ell^{2}} \right)}{\sqrt{1 + \left( - \frac{a_{0}^{4}}{a^{4}} + \frac{a^{2}}{\ell^{2}} \right)}} + 3\pi \frac{a \left(1 + \frac{a^{2}}{l^{2}} \right)}{\sqrt{1 + \frac{a^{2}}{\ell^{2}}}}.
\end{eqnarray}
By re-defining variables
\begin{eqnarray}
z = \left( 1 + \frac{a^{2}}{\ell^{2}} \right)^{3/2},
\end{eqnarray}
the first integration becomes
\begin{eqnarray}
\int_{a_{\mathrm{min}}}^{\infty} \mathcal{F}(a) da = \pi \ell^{2} \int_{z_{\mathrm{min}}}^{\infty} \left( 1 - \frac{1}{\sqrt{1- \frac{a_{0}^{4}/\ell^{4}}{(z-z^{1/3})^{2}}}} \right) dz.
\end{eqnarray}
Then this integration is in a very good form to be calculated numerically. The numerical results as a function of $a_{0}$ are summarized in FIG.~\ref{fig:probability}.

Up to now, we have assumed $V_{0} = 1$ without a loss of generality. However, as we mentioned, after we recover $V_{0}$, the correct subtracted Euclidean action should be $B/V_{0}$, where $B$ is the values obtained in FIG.~\ref{fig:probability}. In this respect, we can summarize the interesting conclusions as follows:
\begin{itemize}
\item[--] If there is no potential and $V_{0} = 1$, then $a_{\mathrm{min}}$ is the only free parameter that characterizes the wormhole. Therefore, this means that \textit{as the size of the throat becomes larger and larger, the probability is exponentially suppressed}.
\item[--] As $V_{0}$ goes to zero, the subtracted action diverges to positive infinity. Therefore, the decay rate approaches to zero. Since $V_{0} \rightarrow 0$ is the Minkowski limit, we can conclude that \textit{the Minkowski space is stable up to fuzzy Euclidean wormholes}, whatever the imaginary part of the scalar field behaves like a ghost field.
\end{itemize}
Regarding the second conclusion, there is one more comment. If we naively calculate the corresponding Euclidean wormhole solution in the asymptotic Minkowski background, the subtracted Euclidean action behaves like $B \sim - a_{0}^{2}$. In this case, we have neglected the term that integrates the cosmological constant over the total volume. If this is true, then the tunneling rate is exponentially enhanced; in addition, this means that the Minkowski space is exponentially unstable, which is contradictory to observations. However, this naive calculation loses the correct action integration of the Minkowski space; this is not that simple, since the vacuum energy is zero but the volume is infinite. This subtle integration can be well defined by adding a cosmological constant. Specifically, by assuming the anti-de Sitter space, the asymptotic classicality of the solution is still maintained (while this does not occur for the asymptotic de Sitter case, and hence it is not suitable). After detailed calculations, what we have shown is that Minkowski is stable up to fuzzy Euclidean wormholes, which is reasonable in terms of simple observations.

\begin{figure}
\begin{center}
\includegraphics[scale=0.85]{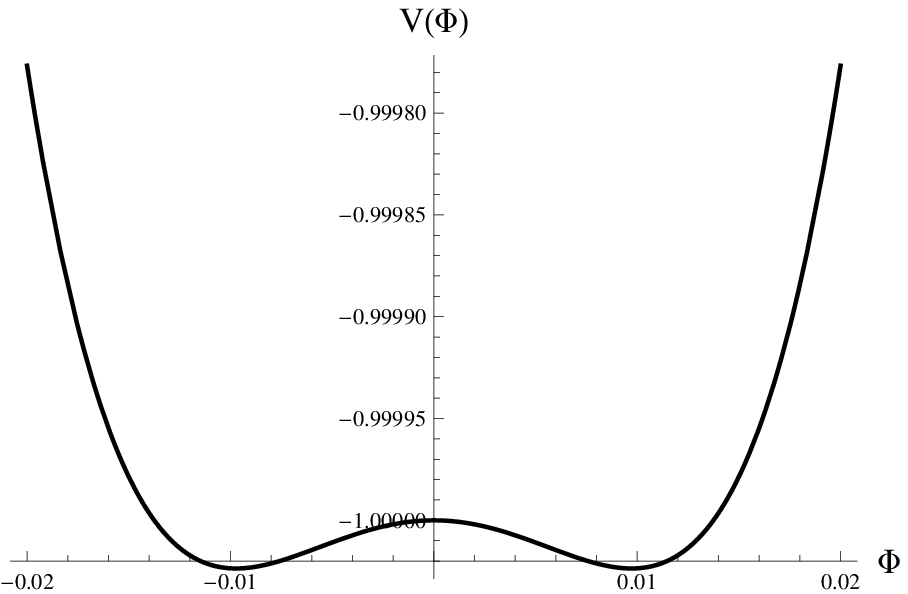}
\includegraphics[scale=0.85]{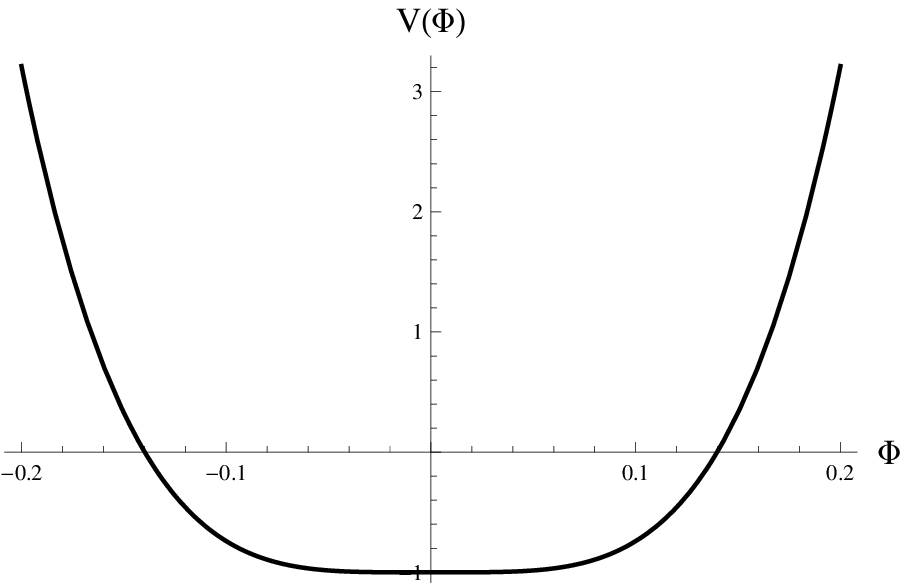}
\caption{\label{fig:pot}Potential $V(\Phi)$ for $\mu^{2} = 1$ and $\lambda = 2645.34485$. The left shows near the symmetric axis, and the right shows the global shape.}
\end{center}
\end{figure}

\subsection{Anti-de Sitter with a potential term}

Now we turn to the potential parameters $\mu^{2}$ and $\lambda$. Although the potential is locally tachyonic, for stability at the local minimum $\Phi = 0$, we assume the Breitenlohner-Freedman bound \cite{Breitenlohner:1982jf}: $\mu^{2}/H_{0}^{2} < 9/4$, where $H_{0} = \sqrt{8\pi/3}$. One example of the potential is given in FIG.~\ref{fig:pot}.

For the large $\tau$ limit, the asymptotic form of the scalar field, in general, should be \cite{Kanno:2012ht}
\begin{eqnarray}
\phi \simeq A_{1} \exp\left[ \left( -\frac{3}{2} H_{0} + \sqrt{\frac{9}{4} H_{0}^{2} - \mu^{2}} \right) \tau \right] + A_{2} \exp\left[ \left( -\frac{3}{2} H_{0} - \sqrt{\frac{9}{4} H_{0}^{2} - \mu^{2}} \right) \tau \right].
\end{eqnarray}
The first term is the slowly-rolling mode and the second term is the quickly-rolling mode. If $A_{1}$ is not zero, then, in general, the first term is dominant. On the other hand, the condition for a finite probability is $A_{1} = 0$ (we will show this soon, see also \cite{Kanno:2012ht}). The choice of $A_{1}$ is related to the choice of the initial condition (of a certain moment) of the field. If the velocity is too large, then the solution will oscillate; if the velocity is too small, then the solution will be dominated by the slowly-rolling mode. In between the two regimes, by choosing a suitable velocity, there is a chance to see the quickly-rolling solution.

\begin{figure}
\begin{center}
\includegraphics[scale=0.85]{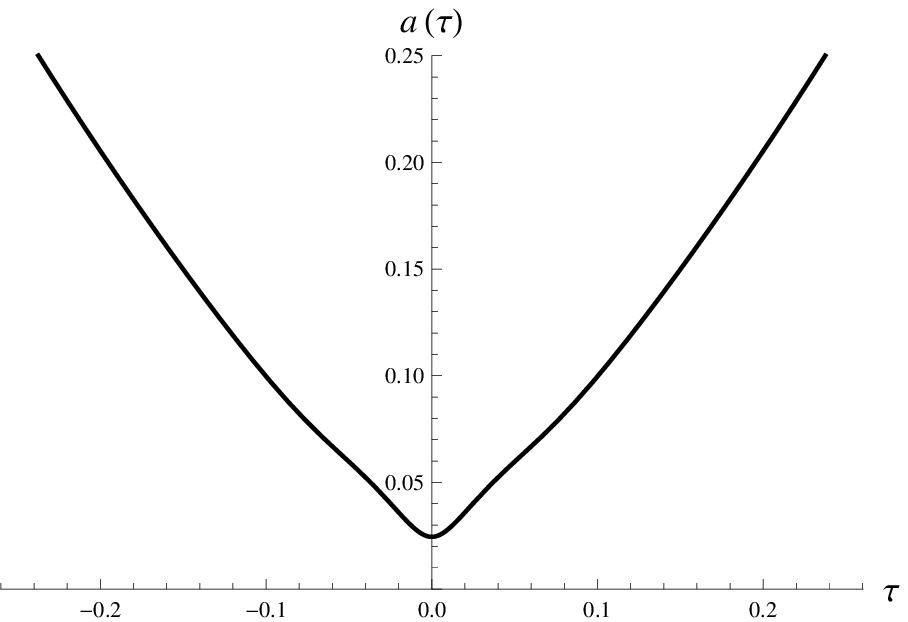}
\includegraphics[scale=0.85]{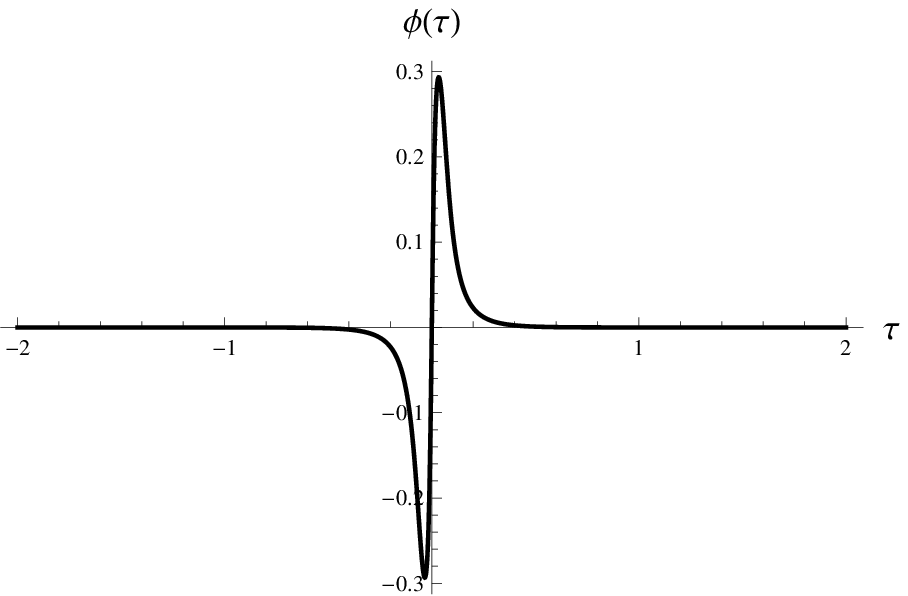}
\caption{\label{fig:sol}$a(\tau)$ and $\phi(\tau) (= -i \Phi)$ for $\dot{\phi}_{0} = 20$.}
\end{center}
\end{figure}

We have no freedom to choose the initial condition. Therefore, in order to tune $A_{1}$, we change the shape of the potential, i.e., choose a suitable $\lambda > 0$. If $\lambda$ is very large, then the scalar field should oscillate around $\phi = 0$. On the other hand, if $\lambda$ is too small, then the solution is dominated by the $A_{1}$ term. Between the two extreme regimes, a chance exists to vanish the $A_{1}$. 

\begin{figure}
\begin{center}
\includegraphics[scale=0.85]{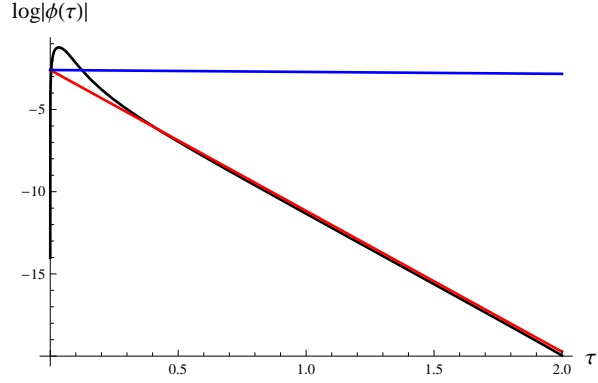}
\caption{\label{fig:logphi}The black curve is $\log |\phi|$, where the blue curve is the slope for the slowly-rolling solution and the red curve is the slope for the quickly-rolling solution. By tuning a suitable $\lambda$, one can obtain a quickly-rolling solution.}
\end{center}
\end{figure}

FIG.~\ref{fig:sol} demonstrates such an example. By tuning $\lambda$, one can obtain an example where the scalar field asymptotically approaches to zero. Therefore, we can ensure the asymptotic classicality. At the same time, by carefully observing the slope of $\log |\phi|$, we can see that it is possible to choose a solution that behaves through the quickly-rolling mode (FIG.~\ref{fig:logphi}).

Then we can see that the subtracted action can be finite, due to the choice of the quickly-rolling mode \cite{Kanno:2012ht}. In the large $\tau$ limit, the solution behaves such that \cite{Lee:2014ufa}
\begin{eqnarray}
\Phi &\simeq& i A_{1,2} \exp\left[ \left( -\frac{3}{2} H_{0} \pm \sqrt{\frac{9}{4} H_{0}^{2} - \mu^{2}} \right) \tau \right],\\
V(\Phi) &=& - 1 - \frac{\mu^{2}}{2} \Phi^{2} - \lambda \Phi^{4},\\
&\simeq& - 1 + \frac{\mu^{2}}{2} A_{1,2}^{2} \exp\left[ \left( -3 H_{0} \pm \sqrt{9 H_{0}^{2} - 4 \mu^{2}} \right) \tau \right] - \lambda A_{1,2}^{4} \exp\left[ \left( -6 H_{0} \pm 2 \sqrt{9 H_{0}^{2} - 4 \mu^{2}} \right) \tau \right],\\
a(\tau) &\simeq& \frac{1}{H_{0}} e^{+H_{0} \tau}.
\end{eqnarray}
Therefore, the Lagrangian becomes
\begin{eqnarray}
4\pi^{2} \left[ a^{3} V - \frac{3}{8\pi} a \right] \simeq \frac{4\pi^{2}}{H_{0}^{3}} \left[ - e^{3 H_{0} \tau} + \frac{\mu^{2}}{2} A_{1,2}^{2} e^{\left( \pm \sqrt{9 H_{0}^{2} - 4 \mu^{2}} \right) \tau} - \lambda A_{1,2}^{4} e^{\left( -3 H_{0} \pm 2 \sqrt{9 H_{0}^{2} - 4 \mu^{2}} \right) \tau} - e^{H_{0} \tau} \right].
\end{eqnarray}
On the other hand, the Lagrangian for the background becomes
\begin{eqnarray}
\frac{4\pi^{2}}{H_{0}^{3}} \left( - e^{3 H_{0} \tilde{\tau}} - e^{H_{0} \tilde{\tau}} \right),
\end{eqnarray}
where this Lagrangian should be integrated by $d\tilde{\tau}$. By matching $\tau = \tilde{\tau}$, the two terms of the integrand cancel out and the only contribution becomes
\begin{eqnarray}
B = \frac{4\pi^{2}}{H_{0}^{3}} \int^{\infty} d\tau \left[ \frac{\mu^{2}}{2} A_{1,2}^{2} e^{\left( \pm \sqrt{9 H_{0}^{2} - 4 \mu^{2}} \right) \tau} - \lambda A_{1,2}^{4} e^{\left( -3 H_{0} \pm 2 \sqrt{9 H_{0}^{2} - 4 \mu^{2}} \right) \tau} \right] + (\mathrm{finite\;numbers}),
\end{eqnarray}
where this integration is finite for the lower sign. This proves that our solution gives a finite result, if it is dominated by the quickly-rolling mode.

In summary, if there is a potential in the scalar field, it is still possible to obtain a fuzzy Euclidean wormhole that is asymptotically classicalized and has non-divergent subtracted action. However, in order to obtain such a solution, we need to carefully tune the shape of the potential. For more detailed calculations of the probability, it is convenient to use the thin-shell approximation, and we provide this calculation in the next section.

\subsection{Analytic continuation to the Lorentzian time}

In previous sections, we discussed the existence and properties of bouncing solutions of $a$. Now we show their physical meanings in terms of Lorentzian spacetime.

We can write more details of the Euclidean metric:
\begin{eqnarray}
ds_{\mathrm{E}}^{2} = d\tau^{2} + a^{2}(\tau) \left( d\chi^{2} + \sin^{2} \chi \; d\Omega_{2}^{2} \right),
\end{eqnarray}
where $0 \leq \chi \leq \pi$ is one of the angle parameters of the three sphere.

In the usual analytic continuation of inhomogeneous instantons, we Wick-rotate following the hypersurface $\chi = \pi/2 + iT$. Then the metric becomes
\begin{eqnarray}\label{eq:lor}
ds^{2} = d\tau^{2} + a^{2}(\tau) \left( - dT^{2} + \cosh^{2} T \; d\Omega_{2}^{2} \right).
\end{eqnarray}
In these coordinates, $0 < T < \infty$ is the time-like parameter (constant $T$ surfaces are space-like) and $\tau$ is the space-like parameter (constant $\tau$ surfaces are time-like). Note that the solution still satisfies the equations of motion up to analytic continuations.

\begin{figure}
\begin{center}
\includegraphics[scale=0.4]{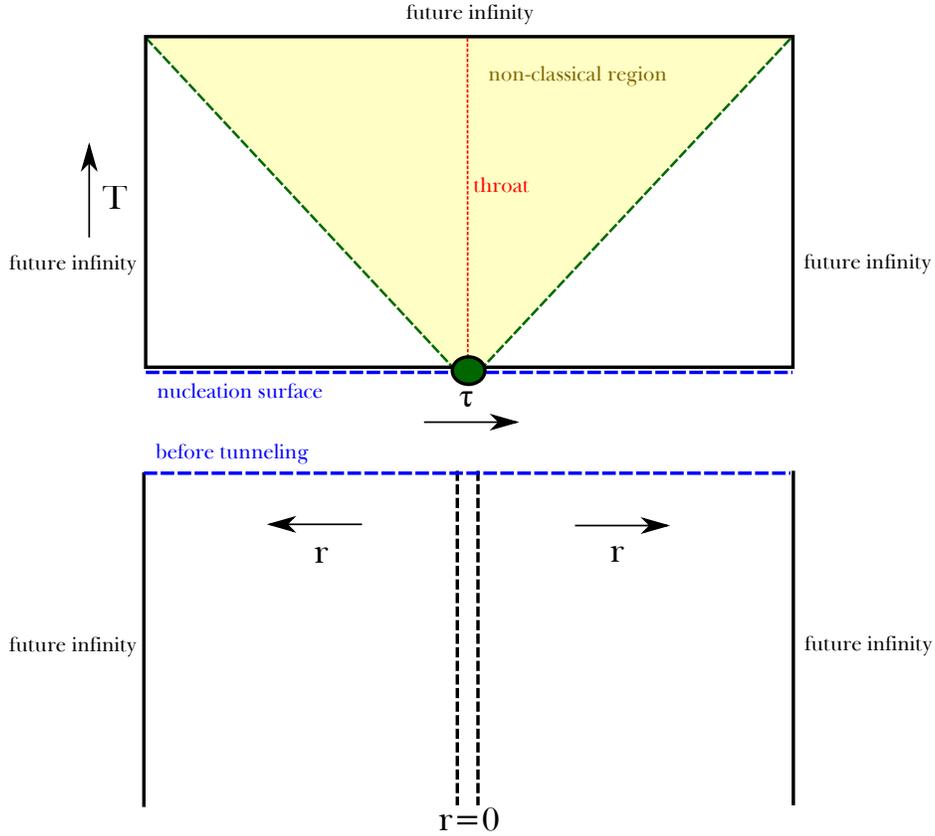}
\caption{\label{fig:causal}The Lorentzian causal structure of the wormhole. The wormhole is nucleated at the blue dashed line at the bottom. At the symmetric point, there is the throat of the wormhole (red dashed line). Non-classical matter (the green colored circle at the bottom) will affect the future and the affected part can be a non-classical region.}
\end{center}
\end{figure}

Note that this Lorentzian metric is a bit different from the usual cases of de Sitter or anti-de Sitter spaces. For pure de Sitter or anti-de Sitter spaces, $a$ can be zero at $\tau = 0$ and this allows that $T \rightarrow \infty$ becomes a null hypersurface \cite{Hawking:1998bn}. In this case, one can define one more Wick-rotation: $T = i\pi/2 + \psi$. However, in Euclidean wormhole cases, $a$ is always non-zero, and, hence, Eq.~(\ref{eq:lor}) is a sufficient analytic continuation.

It is interesting to see the behavior of the throat of the wormhole. At $\tau = 0$, the areal radius of the throat behaves $\sim a_{\mathrm{min}} \cosh T$. Therefore, the throat becomes larger and larger, and as $T$ goes to infinity, the throat eventually becomes infinite. In itself this is not strange, since there is an effective ghost field around the throat, which has been tested numerically \cite{Doroshkevich:2008xm} using double-null simulations \cite{Hong:2008mw}. In addition, the causal future of the throat cannot be classicalized due to the imaginary scalar field. Therefore, by summarizing these observations, we can conclude that, after the Wick-rotation, the non-classical region can reach to the time-like boundaries of the anti-de Sitter space (FIG.~\ref{fig:causal}).

\section{\label{sec:dyn}Dynamics of the throat and further generalizations}

Our original motivation was to obtain solutions that were asymptotically classical. We can obtain asymptotically classical solutions in the \textit{Euclidean sense}, but this is not so trivial in the \textit{Lorentzian sense}. Surely, the causal future of the imaginary part of the scalar field will cover the boundary of the anti-de Sitter spaces as in the green region in FIG.~\ref{fig:causal}. It is not clear whether the effects of the imaginary part of the scalar field are very harmful or nothing special to the boundary.

We cannot completely solve this question in this paper, but we can find a situation where at least the wormhole is hidden by the event horizon of the black hole. Then, in this case, the wormhole throat will not be infinitely expanded (though it could be seen by an asymptotic observer). We can see more details by using the thin-shell approximation.

\subsection{Thin-shell approximation}

If the matter field has only an effect between the two regions, then the thin-shell approximation \cite{Israel:1966rt} will be a good description of this wormhole system \cite{Poisson:1995sv}. We can approximate the left and the right side of the throat as
\begin{eqnarray}
ds_{\pm}^{2} = - f_{\pm}(R)dT^{2} + \frac{1}{f_{\pm}(R)}dR^{2} + R^{2} d\Omega_{2}^{2},
\end{eqnarray}
where $f_{\pm}(R) = 1 - 2M/R + R^{2}/\ell^{2}$ (i.e., the left and right side are symmetric). The induced metric on the shell becomes
\begin{eqnarray}
ds_{\mathrm{shell}}^{2} = - dt^{2} + r^{2} d\Omega_{2}^{2},
\end{eqnarray}
where this $r$ now depends on time. The junction equation between the two regions is \cite{Israel:1966rt}
\begin{eqnarray}
\epsilon_{-} \sqrt{\dot{r}^{2} + f_{-}} - \epsilon_{+} \sqrt{\dot{r}^{2} + f_{+}} = 4 \pi r \sigma,
\end{eqnarray}
where $\epsilon_{\pm}$ denotes the direction of the outward normal direction and $\sigma$ is the tension of the shell, which is a constant if the shell originated from a scalar field. Therefore, we need to choose $\epsilon_{-} = -1$ and $\epsilon_{+} = 1$. This implies that $\sigma < 0$ and the null energy condition on the thin-shell should be violated.

\begin{figure}
\begin{center}
\includegraphics[scale=0.95]{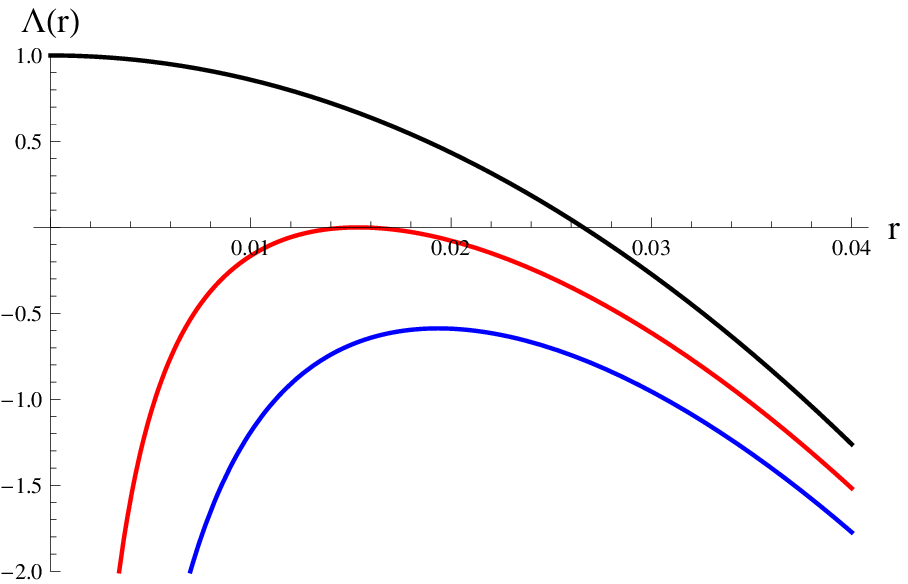}
\includegraphics[scale=0.27]{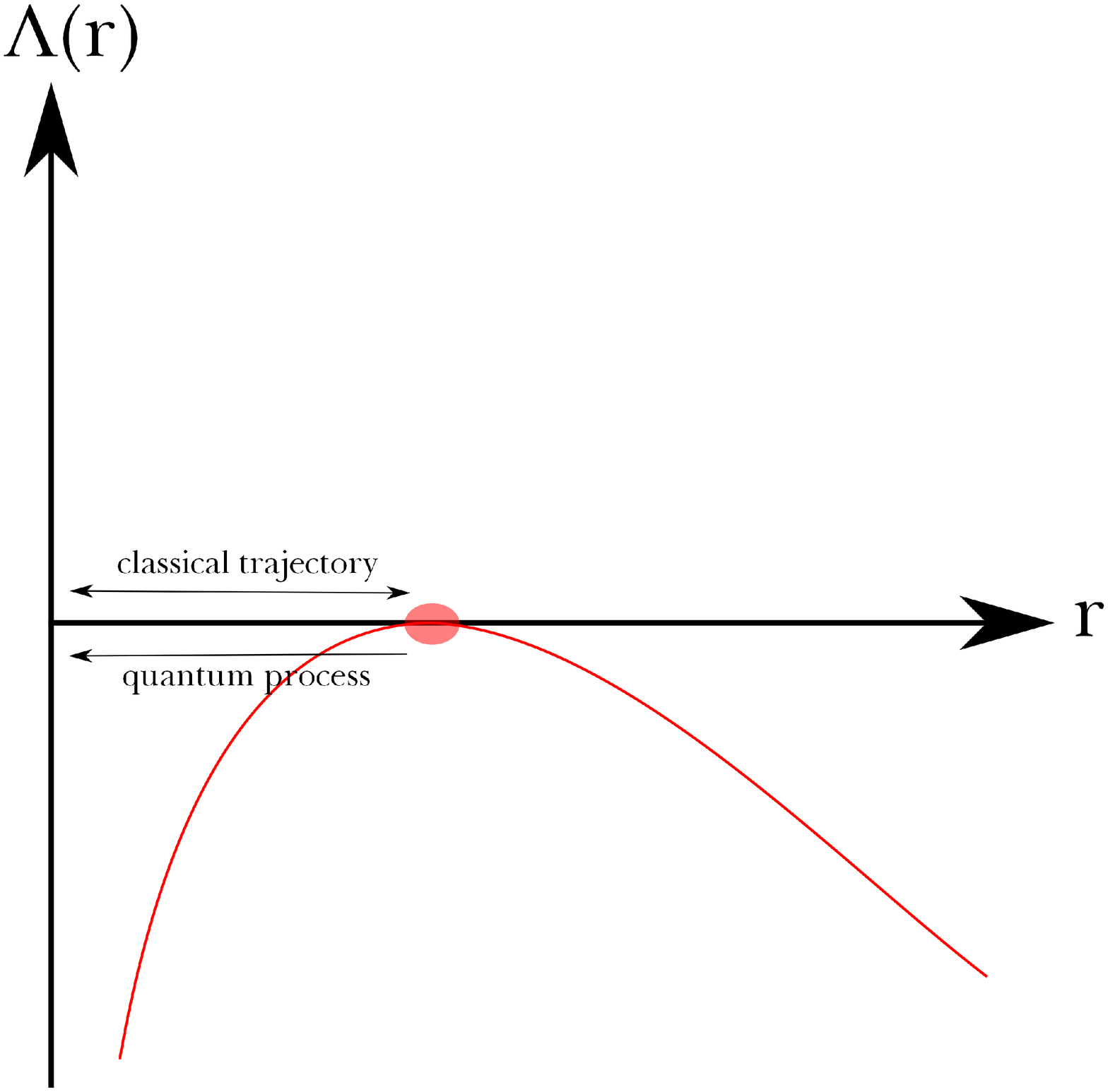}
\caption{\label{fig:effpot}Left: The effective potential $\Lambda(r)$ for $\ell = \sqrt{3/8\pi}$ and $\sigma = -6$ (approximately tuned to fit $a_{\mathrm{min}}$ of the same order of the solution). The black curve is for $M=0$, the red curve is the critical limit $M = M_{\mathrm{cr}}$, and the blue curve is for $M = 2 M_{\mathrm{cr}}$. Right: In the critical limit, one can interpret this several ways; e.g., the wormhole is expanding and contracting, or the wormhole is nucleated at the symmetric point and eventually contracts.}
\end{center}
\end{figure}

Then this equation is simplified by
\begin{eqnarray}
\dot{r}^{2} + \Lambda(r) = 0,
\end{eqnarray}
where
\begin{eqnarray}
\Lambda(r) = 1 - \frac{2M}{r} - \left( 4 \pi^{2} \sigma^{2} - \frac{1}{\ell^{2}} \right) r^{2}.
\end{eqnarray}
Therefore, if $M=0$, as we expected, the wormhole throat expands. By choosing a suitable $\sigma$, we can mimic the dynamics of the wormhole throat in the $O(4)$-symmetric case (FIG.~\ref{fig:effpot}, left).

Using this thin-shell approximation with $M=0$, we can further calculate the tunneling property easily. The subtracted action $B$ is between the solution (wormhole) and the background (two anti-de Sitter spaces), where after the subtraction, the only non-trivial contribution of the solution is on the shell at $a = a_{\mathrm{min}}$ and the only non-trivial contribution of the background is the volume integration for $a \leq a_{\mathrm{min}}$. Therefore,
\begin{eqnarray}
B = \pi \ell^{2} \left[ \left( 1 + \frac{a_{\mathrm{min}}^{2}}{\ell^{2}} \right)^{3/2} - 1 \right] - 2\pi^{2} |\sigma| a_{\mathrm{min}}^{3}.
\end{eqnarray}
By plugging in the condition of $a_{\mathrm{min}}$, we simplify so that
\begin{eqnarray}
\frac{B}{\pi\ell^{2}} = \frac{2\pi|\sigma|\ell}{\sqrt{4\pi^{2}\sigma^{2}\ell^{2}-1}} - 1,
\end{eqnarray}
where this is always positive definite for $2\pi|\sigma|\ell > 1$.

\subsection{Interpretation of generic thin-shell wormholes}

In this thin-shell approximation, we have the freedom to change the asymptotic mass $M$ as well as the tension $\sigma$ as a function of $r$ (see \cite{Poisson:1995sv}; this is possible if the shell originated in a suitable matter field with a specific equation of state, e.g., \cite{Lee:2015rwa}). In this paper, we consider $\sigma$ to have originated from the complexification of the scalar field, and, hence, $\sigma$ should be assumed to be a constant. If we increase $M > 0$, then we lose the $O(4)$-symmetry, but still, it should be physically allowed. Although we will not construct more detailed solutions beyond the thin-shell approximation, it is very reasonable to assume that there should be a corresponding fuzzy instanton, even for the $M>0$ case.

In this regard, the critical mass $M_{\mathrm{cr}} \equiv 1/\sqrt{27 (4\pi^{2}\sigma^{2} - 1/\ell^{2})}$ is an interesting limit (the red curve in FIG.~\ref{fig:effpot}). One can interpret it in several ways.

\begin{figure}
\begin{center}
\includegraphics[scale=0.38]{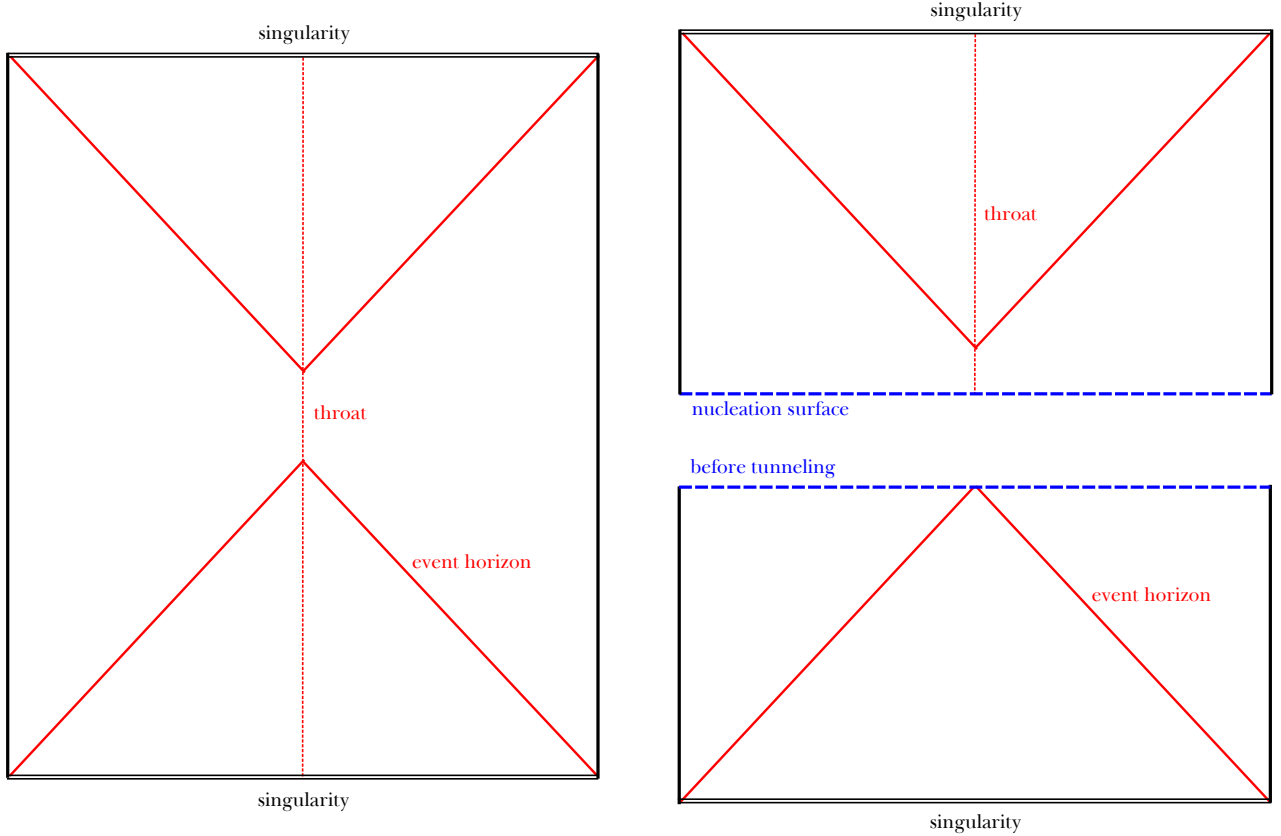}
\caption{\label{fig:causal2}Causal structures of the thin-shell extreme wormholes for the classical trajectory (left) and quantum process (right).}
\end{center}
\end{figure}

For example, in the purely classical way, a wormhole can start from $r=0$ (i.e., an initial singularity\footnote{However, this initial singularity is just due to the time symmetry of the static solution, as in \cite{Blau:1986cw}. This initial condition should be replaced by more physical initial conditions, e.g., the buildable initial condition \cite{Freivogel:2005qh}.}), reach the symmetric point, and collapse to the singularity again. Of course, one can imagine a situation where the wormhole goes over the barrier and expands forever, but, in this section, we will focus only on scenarios in which the wormhole is hidden by the event horizon. In that case, the causal structure will be as shown in the left side of FIG.~\ref{fig:causal2}.

On the other hand, it is possible to think that the wormhole is created at the local maximum of $\Lambda(r)$. This is possible since the local maximum of $\Lambda(r)$ is also a solution of the Euclidean equation of motion of the shell. This is quite similar to the limit of thermal instantons \cite{Gomberoff:2003zh}. In this case, we interpret that the shell begins to collapse after a certain time. Then the final causal structure can be interpreted as shown in the right side of FIG.~\ref{fig:causal2}.

For both interpretations, the throat of the wormhole (and the place where there is an imaginary part of the scalar field) will be naked to the asymptotic observers at infinity. However, it is fair to say that such a wormhole can be hidden eventually inside the event horizon, and, hence, it should not be that severe, as in the case of $M=0$ (as we have seen in FIG.~\ref{fig:causal}). Therefore, as long as we can be sure that the thin-shell versions of the wormholes are included in the class of fuzzy Euclidean wormholes, we will need to carefully consider instantly created wormholes, which cause the Einstein-Rosen bridge to be instantly time-like.

\section{\label{sec:con}Conclusion}

In this paper, we investigated fuzzy Euclidean wormholes in anti-de Sitter space and their analytic continuations to the Lorentzian signatures. We first constructed $O(4)$-symmetric wormholes in detail, and showed that the solutions can give sensible probabilities. We went beyond the $O(4)$-symmetry by using thin-shell approximations in order to see more generic dynamics of the wormhole in the Lorentzian signatures. The $O(4)$-symmetric wormholes should expand as time goes on, but if one considers beyond the $O(4)$-symmetry, they do not necessarily expand, and the wormhole can be hidden by the event horizon.

These Euclidean wormholes are quite interesting but could be potentially harmful to the general framework of the quantum field theory. In particular, we mention two types of problems: the stability issue and the unitarity issue of the anti-de Sitter space.
\begin{itemize}
\item[--] \textit{Quantum stability of the anti-de Sitter space:} In this paper, we considered a tachyonic potential within the Breitenlohner-Freedman bound \cite{Breitenlohner:1982jf}. With a suitable assumption, this space could be stable \cite{Boucher:1984yx}. On the other hand, it was also observed that (by tuning the shape of the potential) space can have a non-perturbative instanton solution that may cause a non-perturbative instability of the background \cite{Kanno:2012ht}. What we can say is that a fuzzy Euclidean wormhole is such an example, and can cause an instability in a local maximum in a non-perturbative way.
\item[--] \textit{Unitarity and the information loss problem:} This violates the cluster decomposition principle \cite{Maldacena:2004rf,ArkaniHamed:2007js} and information from one boundary can be transferred to the other side in the Lorentzian sense. This may cause the loss of information \cite{Hawking:1988ae,Hawking:1976ra}. In addition, if there is a quantum process that makes the Einstein-Rosen bridge instantly time-like, then it will violate the ER=EPR conjecture, since it allows communication between the two time-like boundaries of the anti-de Sitter space \cite{Maldacena:2013xja}; e.g., see \cite{Chen:2016nvj}. One may suggest a tension with some known theorems of the averaged null energy condition \cite{Wall:2011hj}. However, it may still be possible to interpret that: a certain instanton may violate the averaged null energy condition, though if we sum over all instantons, it should not violate the energy condition. Hence, the existence of such a solution itself may not be inconsistent.
\end{itemize}

These conclusions would be very radical, but there may be several loopholes. First, there is a possibility that these fuzzy instantons do not contribute to the path integral (e.g., due to too many negative modes \cite{Coleman:1987rm}). Second, perhaps a more reasonable counter-argument is this: for all our examples, the imaginary part of the scalar field would be seen by the observer at infinity, and this would spoil the classicality of the asymptotic observer. Therefore, if one can assume that the boundary observer should be completely classical, then such a harmful effect from the imaginary scalar field can be a good reason to neglect such a solution in the path integral.

Nonetheless, we are not sure whether such a counter-argument is crucial or not. As we have mentioned in the previous sections, even though the wormhole throat is naked, such an effect would not be too harmful to the asymptotic observer. In that case, they would appear to be subtle issues.

In conclusion, the authors are still very cautious about all the conclusions in this paper. Despite this caution, in any case, the fuzzy Euclidean wormholes that we found in this paper can be included in the wave functions of the Universe. Perhaps this implies that the unitary observer who gathers all paths in the path integral must see some effects of wormholes. Then the unitary observer may see non-classical effects from quantum tunneling. The unitary observer at the boundary may become fully quantum gravitational and the observer may be no more semi-classical \cite{Sasaki:2014spa}. This may reveal the richness of the physics in the anti-de Sitter space \cite{Lee:2014ufa}, and there could be various applications in the context of quantum gravity \cite{Coleman:1988tj} as well as the information loss problem. Can this phenomenon help to explain some of the naked quantum gravitational effects of an evaporating black hole \cite{Hwang:2012nn}? For further detailed interpretations and various subtle problems, we will defer to future works.

\newpage

\section*{Acknowledgments}
DY is supported by Leung Center for Cosmology and Particle Astrophysics (LeCosPA) of National Taiwan University (103R4000). DY is also supported by the Korean Ministry of Education, Science and Technology, Gyeongsangbuk-do and Pohang City for Independent Junior Research Groups at the Asia Pacific Center for Theoretical Physics. SK is supported by IBS under the project code IBS-R018-D1.

\end{document}